\documentstyle [twoside,12pt]{article}
\newcommand{\nc}{\newcommand}
\nc{\la}{\lambda} \nc{\al}{\alpha}
\nc{\th}{\theta}  \nc{\be}{\beta}
\nc{\ga}{\gamma}  \nc{\Ga}{\Gamma}
\nc{\de}{\delta} \nc{\De}{\Delta}
\nc{\si}{\sigma}  \nc{\ka}{\kappa}
\nc{\om}{\omega}  \nc{\te}{\theta}
\nc{\ra}{\rightarrow}
\nc{\beq}{\begin{equation}}
\nc{\eeq}{\end{equation}}
\nc{\beqa}{\begin{eqnarray}}
\nc{\eeqa}{\end{eqnarray}} \nc{\nnb}{\nonumber}
\nc{\dst}{\displaystyle}
\title{ {\bf B.R.S. renormalisation of some on-shell closed algebras of
symmetry transformations :}\\ \Large{the example of supersymmetric non-linear
$\si$ models : 1) the N=1 case} }
\author{Guy Bonneau
\thanks{\noindent Laboratoire de Physique Th\'eorique et des Hautes Energies,
 Unit\'e associ\'ee au CNRS URA 280,~Universit\'e Paris 7,
 2 Place Jussieu, 75251 Paris Cedex 05. Electronic address
bonneau@lpthe.jussieu.fr} }
\date{\today}
\begin{document}
\maketitle

\begin{abstract}
\noindent In order to study in a regularisation free manner the
renormalisability of d=2 supersymmetric non-linear $\si$ models, one has to use
the algebraic BRS methods ; moreover, in the absence of an off-shell
formulation, one has often to deal with open algebras. We then recall in a
pedagogical and non technical manner the standard methods used to handle these
questions and illustrate them on N=1 supersymmetric non-linear $\si$ model in
component fields, giving the first rigorous proof of their renormalisability.
In the special case of compact homogeneous manifolds (non-linear $\si$ model on
a coset space G/H), we obtain the supersymmetric extension of the analysis done
some years ago in the bosonic case.

A further publication will be devoted to extended supersymmetry.
\end{abstract}
\vfill
{\bf PAR/LPTHE/94-10}\hfill  Mars 1994

\newpage
\section{Introduction}
The quantization of extended supersymmetry raises the difficulty of an
``on-shell" formalism. Indeed, if one leaves aside harmonic superspace
\cite{[1]} where firm rules for quantization \footnote{ \  {\sl i.e.} a
subtraction algorithm insuring the locality of the counterterms.} are not at
hand \cite{[a1]}, on the contrary of ordinary superspace \cite{[2]}, one has to
deal with (super)symmetry transformations that are non-linear and closes only
on-shell. This problem was adressed in \cite{[3]} by O. Piguet and K. Sibold
for the Wess-Zumino model as a ``toy-model" and, in a still uncomplete way, by
P. Breitenlohner and D. Maison \cite{[4]} for supersymmetric Yang-Mills in the
Wess-Zumino gauge.

In this series, as new steps, we analyse the d=2, N=1 supersymmetric non-linear
$\si $ model without auxiliary fields and the N=2 and 4 cases in N=1
superfields.

On another hand, and as should be by now well known \cite{[5]}, there is no
consistent regularisation that respects non-linear supersymmetry (dimensional
reduction being mathematically inconsistent \cite{[6]}). Then, if one wants to
settle on a firm basis e.g. the finiteness of N=4 supersymmetric non-linear
$\si$ models, one needs a regulator free treatment : we shall then use the
B.R.S. cohomological methods (\cite{[7]},\cite{[10]}).

The first task is to write a Slavnov identity. In the presence of on-shell
closed algebras, this has first been studied by Kallosh \cite{[k1]}, de Wit and
van Holten \cite{[w1]} and later on systematised by Balatin and Vilkovisky
\cite{[9]} ; as in this work we intend to offer a pedagogical and
self-contained point of view, we prefer to give here a concrete method for the
construction of the effective classical action and of the Slavnov identity, for
a special class of on-shell closed algebras (subsection 2.1). We illustrate
this method with the examples of the d=2, N=1 supersymmetric non-linear $\si $
model without auxiliary fields (subsection 2.2) and the N= 4 case in N=1
superfields (subsection 2.3).

In the same pedagogical and completeness cares, we then recall in section 3 the
essential cohomological tools of the algebraic approach to the
renormalisability proof {\sl \`a la} B.R.S., and explain with some details the
power of the  ``filtration" method for the cohomological analyses of highly
non-linear nihilpotent linearized Slavnov operators $S_L$ (the so-called
spectral sequence method \cite {[13]}). Some grading being introduced
(``filtration" operator), $S_L^0$, the lowest order of the Slavnov operator is
still nihilpotent and its cohomology is simpler to find. In subsection 3.2, we
then skech the proof of the ``filtration" theorem which asserts the isomorphism
between this cohomology space and the one of the full Slavnov operator, in the
special case where the cohomology of $S_L^0$ is empty in the Faddeev-Popov
charged sectors \cite{[7]}. An appendix gives a sketch of the modifications of
the ``filtration" theorem when the cohomology of $S_L^0$ is not empty in the
Faddeev-Popov charged sectors (\cite{[7]},\cite{[b10]}).

The method is then examplified in section 4 on the d=2, N=1 supersymmetric
non-linear $\si $ model in component fields (without auxiliary fields) leaving
the  N= 4 case in N=1 superfields (and as an intermediary step the  N= 2
supersymmetry) to next publications \cite{[12]}. Another interesting case would
be super Yang-Mills theories in 4 space-time dimensions.

Of course, here we are mainly interested in the renormalisation of the
supersymmetry transformations : as discussed by Friedan \cite{[8]}, the action
of a non-linear $\si$ model may be identified with a distance on a Riemannian
manifold $\cal M$, the metric depending {\sl a priori} on an infinite number of
parameters. One then speaks of ``renormalisability in the space of metrics" or
``{\sl \`a la } Friedan". When there exist extra  isometries, for example in
the case of the non-linear $\si$ models on coset spaces (homogeneous
manifolds), the number of such physical parameters becomes finite and in ref.
\cite{[7]} we have proved the U.V. renormalisability of these isometries in the
purely bosonic case. The present work gives the necessary supersymmetric
extension in section 5. On the other hand, in the absence of isometries, our
aim is the proof that no extra difficulty occurs in the supersymmetric
extension of these non-linear $\si$ models. We shall then prove in section 4
that the cohomology of $S_L^0$ is empty in the Faddeev-Popov charged sectors,
which ensures the renormalisability of the theory `` in the space of metrics".

\section{Slavnov identity for on-shell algebras}
\subsection{General method}
Let $\Phi ^{a}(x)$ and $A[\Phi]$ respectively be a set of fields (bosonic or
fermionic) and the classical action of a theory. We want to analyse a global,
non-linear symmetry transformation
\beq\label{a1}
\de \Phi^{a}(x) = \epsilon^i W_i \Phi^{a}(x)
\eeq
whose generators satisfy the following (anti)commutation relations :
\beq\label{a2}
[W_i ,W_j ]\Phi^{a}(x)= f_{ij}^k W_k\Phi^{a}(x) + X_{ij}^{ab} \frac{\de
A[\Phi]}{\de \Phi^{b}(x)}
\eeq
and where $X_{ij}^{ab}$ is $ a \; priori$ a field dependent quantity \footnote{
\  To fix the notations, we consider bosonic fields and transformations : then
$X_{ij}^{ab}$ will be antisymmetric in (i,j). Notice also that the integration
over x of (\ref{a2}) multiplied by $\frac{\de A[\Phi]}{\de\Phi^{a}(x)}$ shows
that if $X_{ij}^{ab}$ was symmetric in (a,b), the algebra (\ref{a2}) could be
rewritten as a closed one, with field dependent structure constants, a case
that we exclude from our analysis : therefore, in the following, we restrict
ourselves to the cases where $X_{ij}^{ab}$ will be antisymmetric with respect
to (a,b). \newline The necessary modifications for e.g. supersymmetry
transformations are evident. }. This situation is a special case among the ones
studied by Balatin and Vilkovisky \cite{[9]}, but, as we want to be as much
pedagogical as possible, we prefer to examplify the method on a particular
class of on-shell algebras.
Of course, on mass-shell, the ``algebra" (\ref{a2})  takes a canonical form.

As usual \cite{[10]}, in view of quantization one takes as classical effective
action
\beq\label{a3}
\Ga^{class.} = A + \int dx\, \ga_{a}(x)[W_i\Phi^{a}(x)]C^i
\eeq
where $C^i$ and $\ga_{a}(x)$ are respectively constant anticommuting
Faddeev-Popov parameters and external anticommuting \footnote{\  See the last
line of footnote 2.} sources for the B.R.S. variation of the fields
$\Phi^{a}(x)$.

\noindent Then, although the $W_i$'s transformations are generally non-linear,
the variation of $\Ga^{class.} $ will produce the commutator of two
transformations and then will be expressible as a functional derivative with
respect to the source $\ga_a$. More precisely :
\beqa\label{a4}
S \Ga &\equiv & \int dx \frac{\de \Ga}{\de \ga_{a}(x)}\frac{\de \Ga}{\de
\Phi^{a}(x)} \nnb \\
&=& \int dx [W_i\Phi^{a}(x)]C^i\frac{\de A}{\de \Phi^{a}(x)} +  \int dx
[W_i\Phi^{a}(x)]C^i\frac{\de \left[ \int dy
\ga_{b}(y)W_j\Phi^{b}(y)\right]}{\de  \Phi^{a}(x)} C^j \nnb  \\
= \de A &+& {1\over2} C^iC^j \left[ f_{ij}^{k}\frac{\de \Ga}{\de C^k}
 -  \int dx \ga_{b}(x)X_{ij}^{bc}\left( \frac{\de \Ga}{\de \Phi^{c}(x)} -
\frac{\de \left[\int dy \ga_{a}(y)W_k\Phi^{a}(y)\right]}{\de \Phi^{c}(x)} C^k
\right)\right]
\eeqa
At this point we suppose the invariance of the classical action $ A=
A^{inv.}$. Then :
$$ S\Ga - {1\over2} C^iC^j f_{ij}^{k}\frac{\de \Ga}{\de C^k} = -{1\over2}
C^iC^j \int dx X_{ij}^{ab} \ga_{a}(x) \frac{\de \Ga}{\de \Phi^{b}(x)} + {\rm
[3-ghosts\ terms\ ]}\ .$$
In order to suppress the 2-ghosts terms, we modify the classical action
(\ref{a3})
\beq\label{a5}
\Ga^{tot.} = A^{inv.} + \int dx\, \ga_{a}(x)[W_i\Phi^{a}(x)]C^i - {1\over4}
\int dx\, \ga_{a}(x)X_{ij}^{ab}\ga_{b}(x)C^i C^j .
\eeq
Then \footnote{ \  $(G[\Phi(x)]),_c$ means $\frac{\de}{\de \Phi^{c}(x)}\int
dy(G[\Phi(y)]) $}
\beqa\label{a6}
S \Ga^{tot.} - {1\over2} C^iC^j f_{ij}^{k}\frac{\de \Ga^{tot.}}{\de C^k} = -
{1\over2} C^iC^j  \int dx\left[ X_{ij}^{ab} + X_{ij}^{ba}\right] \ga_{b}(x)
\frac{\de \Ga^{tot.}}{\de \Phi^{a}(x)} \nnb \\
 + {1\over2} C^iC^jC^k  \int dx \left[ X_{ij}^{bc} \ga_{b}(x) \left(\ga_{a}(x)
W_k \Phi^{a}(x) \right),_c - {1\over2}(- f_{ij}^{l} X_{lk}^{ba} + W_k
X_{ij}^{ba})  \ga_{b}(x)\ga_{a}(x)\right] \nnb \\
+ {1\over8} C^iC^jC^kC^l \int dx \ga_{b}(x) X_{ij}^{ab} \left(
X_{kl}^{cd}\ga_{c}(x)\ga_{d}(x)\right),_a
\eeqa
The 2-ghosts terms cancell due to the antisymmetry of $ X_{ij}^{ab} $ in the
interchange a $ \leftrightarrow$ b . The 3-ghosts ones will be analysed through
the Jacobi identity  $$C^iC^jC^k [W_k,[W_i,W_j]] = 0  $$ giving on-shell :
\beq\label{a7}
C^iC^jC^k f_{ij}^{l}f_{lkn} = 0
\eeq
and then :
\beq\label{a8}
C^iC^jC^k \left(\left[-f_{ij}^{l} X_{lk}^{ba} + W_k X_{ij}^{ba} -  X_{ij}^{ca}
\left( W_k \Phi^{b}(x) \right),_c \right]\frac{\de A}{\de \Phi^{a}(x)} +
X_{ij}^{ba} W_{k}\left(\frac{\de A}{\de \Phi^{a}(x)}\right) \right) = 0
\eeq
The invariance of the classical action is then used to transform equation
(\ref{a8}) into :
\beq\label{a99}
C^iC^jC^k\left(\left[- f_{ij}^{l} X_{lk}^{ba} + W_k X_{ij}^{ba} -  X_{ij}^{ca}(
W_k \Phi^{b}(x)),_c \right]\frac{\de A^{inv.}}{\de \Phi^{a}(x)} -
X_{ij}^{bc}\int dy \frac{\de ( W_k \Phi^{a}(y))}{\de \Phi^{c}(x)} \frac{\de
A^{inv.}}{\de \Phi^{a}(y)}\right) = 0.
\eeq
Notice that when the transformation $ W_{k}(\Phi^{a}(y))$ involves only the
fields and not their derivatives, equation (\ref{a99}) may be ``divided" by
$\frac{\de A^{inv.}}{\de \Phi^{a}(x)}$ and, as a consequence, the 3-ghosts term
of (\ref{a6}) does vanish.

This gives the spirit of the construction of the Slavnov identity for on-shell
closed algebras; in the generic case, the total action (\ref{a5}) has to be
modified by addition of 3-[and even more] ghosts terms in order to obtain from
equation (\ref{a99}) [and similarly obtained ones at higher order in the number
of ghosts] the looked-for vanishing of the 3-[and higher order] ghosts terms
and the final Slavnov identity :
\beq\label{a10}
S \Ga^{tot.} \equiv  \int dx \frac{\de \Ga^{tot.}}{\de \ga_{a}(x)}\frac{\de
\Ga^{tot.}}{\de \Phi^{a}(x)}  = {1\over2} C^iC^j f_{ij}^{k}\frac{\de
\Ga^{tot.}}{\de C^k} \;\; .
\eeq
In the present work, we shall show that in some interesting examples this is
not necessary and that equation (\ref{a10}) will hold true with a  $\Ga^{tot.}$
given by (\ref{a5}). Moreover, due to the algebra (\ref{a2}), one expects
nihilpotency for the linearized Slavnov operator $S_L$ defined through :
$$ S(\Ga + \epsilon \Ga^1) \equiv S \Ga + \epsilon S_L \Ga^1 + {\cal O}
(\epsilon^2) \ . $$

\subsection{N=1 supersymmetric non-linear $\si$ models in component fields}

Let us examplify this method on d=2, N=1 supersymmetric non-linear $\si$ models
in component fields {$\phi^{i} (x), \psi^{j}_{\pm}(x)$} (i,j,.. = 1,2,..n). In
light-cone coordinates and {\sl in the absence of torsion}, the non-linear N=1
supersymmetry transformations and the invariant action respectively write
\beqa\label{B1}
\de\phi^i &=& \epsilon^{+}\psi^{i}_{+} +  \epsilon^{-}\psi^{i}_{-} \nnb \\
\de\psi^{i}_{\pm} &=& i \epsilon^{\pm}\partial_{\pm}\phi^{i} +
\epsilon^{\mp}\Ga^i_{jk}\psi^{j}_{\pm} \psi^{k}_{\mp}
\eeqa
\beq\label{B11}
 A^{inv.} = \int d^2 x \{ g_{ij}(\phi)[\partial_{+}\phi^{i}\partial_{-}\phi^{j}
+ i(\psi^{i}_{+}D_{-}\psi^{j}_{+} + \psi^{i}_{-}D_{+}\psi^{j}_{-} )] +
{1\over2}R_{ijkl}\psi^{i}_{+}\psi^{j}_{+}\psi^{k}_{-}\psi^{l}_{-} \}
\eeq
where the covariant derivatives are
 $$ D_{\pm}\psi^{i}_{\mp} = \partial_{\pm}\psi^{i}_{\mp}
+\Ga^i_{jk}\partial_{\pm}\phi^{j}\psi^{k}_{\mp}$$ and where $\Ga^i_{jk}$ and
$R_{ijkl}$ are respectively the (symmetric) connexion and Rieman tensor
associated to the target space metric $g_{ij}[\phi]$. The properties of the
theory will be more transparent in tangent space where the metric is the
Khr\"onecker $\de_{ab}$, the spin connexion $\om ^a_{bc} $, the Riemann tensor
$ R_{abcd}$ and the tangent space fermions $\psi^{a}_{\pm}$ are related to
world space ones  $\psi^{i}_{\pm}$ through the vielbeins $e_{i}^{a}[\phi]$ :
$$ g_{ij} = \de_{ab}e_{i}^{a}e_{j}^{b} \ \ ; \ \psi^{a}_{\pm} = \psi^{i}_{\pm}
e_{i}^{a} $$
$$\om ^a_{bc} = e_{i}^{a}e^{j}_{b}e^{k}_{c}\Ga^i_{jk} -  e^{j}_{b}e^{i}_{c}
\partial_j  e_{i}^{a} \ \ ; \ \ R_{abcd} =
R_{ijkl}e^{i}_{a}e^{j}_{b}e^{k}_{c}e^{l}_{d} \ \ {\rm where \ \ }
e_{i}^{a}e^{i}_{b} = \de^{a}_{b}\ . $$
With $ e^{a}_{\pm} = e_{i}^{a}\partial _{\pm}\phi^i$, the covariant derivatives
and the invariant action are respectively :
\beq\label{B3}
D_{\pm}\psi^{a}_{\mp} = \partial_{\pm}\psi^{a}_{\mp} +
\om^a_{bc}e^{b}_{\pm}\psi^{c}_{\mp}
\eeq
\beq\label{ B33}
A^{inv.} = \int d^2 x \{ \de_{ab}[ e^{a}_{+}e^{b}_{-} +
i(\psi^{a}_{+}D_{-}\psi^{b}_{+} + \psi^{a}_{-}D_{+}\psi^{b}_{-} )] +
{1\over2}R_{abcd}\psi^{a}_{+}\psi^{b}_{+}\psi^{c}_{-}\psi^{d}_{-} \}\ .
\eeq
Then, the (highly non-linear) supersymmetry transformations are :
\beqa\label{B4}
\de\phi^i &=&  e^{i}_{a}( \epsilon^{+}\psi^{a}_{+} +  \epsilon^{-}\psi^{a}_{-})
\nnb \\
\de\psi^{a}_{\pm} &=& i \epsilon^{\pm} e^{a}_{\pm} -
\om^a_{bc}\psi^{c}_{\pm}(\epsilon^{+}\psi^{b}_{+} + \epsilon^{-}\psi^{b}_{-})
\eeqa
and the algebra of equ.(\ref{a2}) specifies to :
\beqa\label{B5}
\{ W_{\pm},W_{\pm} \}\phi^i &=& 2i\partial_{\pm}\phi^i \nnb \\
\{ W_{+},W_{-} \}\phi^i &=& 0 \nnb \\
\{ W_{\pm},W_{\pm} \}\psi^{a}_{\pm} &=&  2i\partial_{\pm}\psi^{a}_{\pm} \nnb \\
\{ W_{\pm},W_{\pm} \}\psi^{a}_{\mp} &=&  2i\partial_{\pm}\psi^{a}_{\mp} -
\de^{ab}\frac{\de A^{inv.}}{\de \psi^{b}_{\mp}} \nnb \\
\{ W_{+},W_{-} \}\psi^{a}_{\pm} &=&  {1\over2} \de^{ab}\frac{\de  A^{inv.}}{\de
\psi^{b}_{\mp}}\;\; .
\eeqa
As one is only concerned by integrated local functionals (i.e. trivially
translation invariant ones), we forget about the linear translation operators
$P_\pm \equiv  i\partial_\pm$, to which anticommuting Faddeev-Popov parameters
$p^\pm $ should be associated, and do not add in $\Ga^{class.}$ of
equ.(\ref{a3}) the effect of translations on the fields $\phi^i$ and
$\psi^{a}_{\pm}$ . Then the total effective action of equ.(\ref{a5}) writes:
\beqa\label{B6}
\Ga^{tot.} & = & A^{inv.} + \int d^2 x \{\ga_{a}^{+}(x)[iC^+ e^{a}_{+} +
\om^a_{bc}\psi^{c}_{+}(C^+\psi^{b}_{+} + C^-\psi^{b}_{-})]\nnb \\
& + & \ga_{a}^{-}(x)[iC^- e^{a}_{-} + \om^a_{bc}\psi^{c}_{-}(C^+\psi^{b}_{+} +
C^-\psi^{b}_{-})] + \eta_i(x)[ C^+ e^{i}_{a}\psi^{a}_{+} + C^-
e^{i}_{a}\psi^{a}_{-}]\} \nnb \\
& + & {1\over4}\int d^2 x \de^{ab}\{\ga_{a}^{+}(x)\ga_{b}^{+}(x)(C^-)^2
+\ga_{a}^{-}(x)\ga_{b}^{-}(x)(C^+)^2 - 2\ga_{a}^{+}(x)\ga_{b}^{-}(x)(C^+ C^-)\}
\eeqa
where $C^{\pm}$, $\ga_{a}^{\pm}(x)$ and $\eta_i(x)$ are respectively commuting
Faddeev-Popov parameters,  commuting fermionic and  anticommuting bosonic
sources. Due to the simplicity of the algebra (\ref{B5})\footnote{\ Notice that
in this N=1 supersymmetric case, an off-shell formalism in fact does exist.},
the Slavnov identity of equ.(\ref{a10}) holds and writes :
\beqa\label{B7}
S \Ga^{tot.}&\equiv &\int d^2 x \{ \frac{\de \Ga^{tot.}}{\de \ga_a^{+}}
\frac{\de \Ga^{tot.}}{\de \psi^a_{+}} + \frac{\de \Ga^{tot.}}{\de \ga_a^{-}}
\frac{\de \Ga^{tot.}}{\de \psi^a_{-}} + \frac{\de \Ga^{tot.}}{\de \eta_i}
\frac{\de \Ga^{tot.}}{\de \phi^i} \} \nnb \\
&=& \int d^2 x\{(C^+)^2 ( \eta_k i\partial_+\phi^k + \ga_a^+
i\partial_+\psi^a_+) + (C^-)^2 ( \eta_k i\partial_-\phi^k + \ga_a^-
i\partial_-\psi^a_-)\} \ .
\eeqa
Moreover, one can also check that the linearized Slavnov operator :
\beqa\label{B8}
S_L \equiv \int & d^2 x & \{ \left( \frac{\de \Ga^{tot.}}{\de \ga_a^{+}}\right)
\frac{\de}{\de \psi^a_{+}} + \left( \frac{\de \Ga^{tot.}}{\de \ga_a^{-}}\right)
\frac{\de}{\de \psi^a_{-}} + \left( \frac{\de \Ga^{tot.}}{\de
\psi^a_{+}}\right) \frac{\de}{\de \ga_a^{+}}
+ \left(\frac{\de \Ga^{tot.}}{\de \psi^a_{-}}\right) \frac{\de}{\de \ga_a^{-}}
\nnb \\
& + &  \left(  \frac{\de \Ga^{tot.}}{\de \eta_i}\right) \frac{\de}{\de \phi^i}
+ \left(\frac{\de \Ga^{tot.}}{\de \phi^i}\right) \frac{\de}{\de \eta_i} \}
\eeqa
is nihilpotent : $(S_L)^2 = 0$, when acting in the space of integrated local
functionals.
The quantization of this theory will be studied in section 4.

\subsection{N=4 supersymmetric non-linear $\si$ models in N=1 superfields}

Consider now d=2, N=4 supersymmetric non-linear $\si$ models in N=1 superfields
$\Phi^{i} (x,\te)$ (i, j,.. = 1,2,..4n). In light-cone coordinates and in the
absence of torsion, the non-linear N=4 supersymmetry transformations and the
invariant action respectively write :
\beqa\label{C1}
\de\Phi^i &=& J_{a\,j}^{i}(\Phi)[\epsilon^{+}_a D_{+}\Phi^j + \epsilon^{-}_a
D_{-}\Phi^j ] \ \ ,\ \ a = 1, 2, 3. \nnb\\
A^{inv.} &=& \int d^2 x d^2\te  g_{ij}(\Phi)[D_{+}\Phi^{i}D_{-}\Phi^{j}]
\eeqa
where the covariant derivatives $ D_{\pm} = \frac{\partial}{\partial\te^{\pm}}
+ i \te^{\pm}\partial_{\pm} $ satisfy
\beq\label{C2}
\{ D_{\pm},D_{\pm} \} = 2i\partial_{\pm} \ \ \ \ \{ D_{+},D_{-} \} = 0\ \ .
\eeq
As is well known (see for example ref.\cite{[11]}), N=4 supersymmetry needs the
 $
J_{a\,j}^{i}(\Phi)$ to be a set\footnote{\ As a matter of facts, it is
sufficient to have 2 anticommuting complex structures : then the product
$J_{3\,k}^{i} \equiv J_{1\,j}^{i} J_{2\,k}^{j}$ offers a third complex
structure \ .} of anticommuting integrable complex structures according to :
$$ J_{a\,j}^{i}(\Phi) J_{b\,k}^{j}(\Phi) = -\de_{ab}\de^i_k +
\epsilon_{abc}J_{c\,k}^{i}(\Phi) \ ,$$ and the invariance of the action needs
the target space to be hyperk\"{a}hler :

$\ast$ the metric is hermitic with respect to each complex structure
$$J_{a}^{ij} \equiv J_{a\,k}^{i}g^{kj} = - J_{a}^{ji}$$

$\ast$ the $J_{a\,j}^{i}$ are covariantly constant with respect to the metric
$g_{ij}$
$$D_{k}J_{a\,j}^{i} \equiv \partial_k J_{a\,j}^{i} +\Ga^i_{kl}J_{a\,j}^{l} -
\Ga^l_{kj}J_{a\,l}^{i} = 0 $$
where $\Ga^n_{kl}[g_{ij}]$ is the ( symmetric) connexion.

Then the algebra of equ.(\ref{a2}) specifies to :
\beqa\label{C5}
\{ W_{a\pm},W_{b\pm} \}\Phi^i &=& 2i\de_{ab}\partial_{\pm}\Phi^i \nnb \\
\{ W_{a+},W_{b-} \}\Phi^i &=& \epsilon_{abc}J_{c}^{ij} \frac{\de A^{inv.}}{\de
\Phi^j}
\eeqa
{\sl i.e.} $ X_{ij}^{ab} \equiv 2\epsilon_{abc}J_{c}^{ij}\ .$ As in the
previous section, we forget about the linear translation operators $P_\pm
\equiv  i\partial_\pm $, to which anticommuting Faddeev-Popov parameters $p^\pm
$ should be associated, and do not add in $\Ga^{class.}$ of equ.(\ref{a3}) the
effect of translations on the fields $\Phi^i$ .
Then the total effective action of equ.(\ref{a5}) writes \footnote{\ As
previously mentionned, we consider only torsionless metrics and, as a
consequence, there is a ``parity" invariance :
in the interchange ($ + \leftrightarrow - $), $ d^2 \te$ and $\eta_i $ get a
minus sign wether $\Phi^i$ is unchanged. Under this hypothesis,  {\sl there
will be no room for a chiral anomaly}.} :
\beq\label{C6}
\Ga^{tot.} = A^{inv.} + \int d^2 x d^2\te \{ \eta_i J_{a\,j}^{i}(\Phi)[
d_a^{+}D_{+}\Phi^j + d_a^{-}D_{-}\Phi^j ] - {1\over2}\epsilon_{abc}\eta_i\eta_j
J_{c}^{ij}(\Phi)d_a^{+}d_b^{-} \}
\eeq
where $d_a^{\pm}$ and $\eta_i(x)$ are respectively commuting  Faddeev-Popov
parameters and  anticommuting bosonic sources. Here also, the Slavnov identity
of equ.(\ref{a10}) holds and writes :
\beq\label{C7}
S \Ga^{tot.} \equiv \int d^2 x d^2\te \frac{\de \Ga^{tot.}}{\de \eta_k}
\frac{\de \Ga^{tot.}}{\de \Phi^k} = \int d^2 x d^2\te[(d_a^+)^2 ( \eta_k
i\partial_+\Phi^k)  + (d_a^-)^2 ( \eta_k i\partial_-\Phi^k) ]\ .
\eeq
Moreover, one can also check that the linearized Slavnov operator :
\beq\label{C8}
S_L \equiv \int d^2 x d^2\te\{ \frac{\de \Ga^{tot.}}{\de \eta_k} \frac{\de}{\de
\Phi^k} + \frac{\de \Ga^{tot.}}{\de \Phi^k} \frac{\de}{\de \eta_k} \}
\eeq
is nihilpotent : $(S_L)^2 = 0$, when acting in the space of integrated local
functionals.
These are not trivial results as in that N = 4 case, no finite set of auxiliary
fields exists.

Notice that in the N=2 case - where there is only one complex structure -,
there is no bilinear terms in the sources $\eta_i$ in equation (\ref{C6}) : as
a matter of fact, in that case the supersymmetry algebra on N=1 superfields
closes off-shell.

The quantization of these theories will be studied in a second paper of this
series \cite{[12]}.

\section{Agebraic approach to the renormalisability proof}
\subsection{Generalities}

As recalled in refs.(\cite{[7]},\cite{[5]}), the renormalisation program
consists in solving two main problems. These are :

i) all possible breakings which might affect the Ward identities, order by
order in the radiative corrections, should be reabsorbed by a suitable choice
of (``finite") counterterms [absence of anomalies in the theory],

ii) Ward identities being so ensured at a given order, the (``infinite")
counterterms needed for the finiteness of the renormalized Green functions
should be uniquely identified (up to a field redefinition) by the parameters
characterizing the classical action [stability condition].

If the Ward, or rather Slavnov, identity writes
\beq\label{D1}
S\Ga = 0 \;(\rm or\; a\; classical\; quantity\; as \; in\;
equs.(\ref{B7},\ref{C7})\;)
\eeq
the Quantum Action Principle ensures that, up to the first non-trivial order,
the breaking of the Slavnov identity (\ref{D1}) corresponds to the insertion of
a local functional of Faddeev-Popov charge +1, $\De_{[+1]}$ :
$$ S\Ga = (\hbar)^p\De_{[+1]} + {\rm higher\ orders} $$
and $S_L$ being the linearized, nihilpotent Slavnov operator, one gets
$$S_L\De_{[+1]} = 0\ .$$
Any trivial cohomology $\De_{[+1]} = S_L\De_{[0]}$ corresponds to the
looked-for ``finite" counterterms as
$$S(\Ga - (\hbar)^p\De_{[0]}) = 0 + {\rm higher\ orders} \ .$$
In other words, this means that point i) supposes the vanishing of the
cohomology of $S_L$ in the  Faddeev-Popov charge one sector.

\noindent As regards the stability condition
$$ S(\Ga^{class.} +\hbar \Ga_{[0]}) = 0\ ,$$
$\Ga_{[0]}$, a pertubation of the classical action, is a Faddeev-Popov neutral
local functional which must satisfy :
$$ S_L\Ga_{[0]} = 0\ . $$
Any trivial cohomology $\De_{[0]} = S_L\De_{[-1]}$ may be shown to correspond
to non physical field and source redefinitions. In other words, this means that
point ii) supposes that the dimensionality of the cohomology space of this
$S_L$ operator in the neutral  Faddeev-Popov sector is equal to the number of
``physical" parameters of the classical action.
$$ $$

In the presence of highly non-linear Slavnov operators such as those of
equations (\ref{B8},\ref{C8}), it is technically useful to ``aproximate" the
complete $S_L$ operator by a simpler one $S_L^{(0)}$ through a suitably chosen
``filtration"(counting operation)\cite{[13]} such as the cohomology spaces of
$S_L$ and $S_L^{(0)}$ are isomorphic. This relies on a theorem proved in
ref.\cite{[7]} which asserts that :

\noindent {\bf If the cohomology of $S_L^{(0)}$ is trivial in the
Faddeev-Popov charged sectors, then the same is true for $S_L$ and their
cohomology spaces in the neutral sector are isomorphic.}

In order that this work be as much self-contained as possible, we sketch the
proof in the next subsection, leaving to the appendix its extension to the case
where  $S_L^{(0)}$ has some non-trivial cohomology in the Faddeev-Popov charged
sectors, a situation that will occur in the N=2 and 4 cases \cite{[12]}.

\subsection{The filtration theorem }

Let $S_L$ be a nihilpotent operator which acts in the linear space V of
translation invariant functionals \footnote{\ The analysis is made even easier
when one can adapt the formalism to ``local" cohomology where the linear space
is a space of functions of the fields, sources and their derivatives, taken as
independent variables. In such a case, a Fock space formulation is at hand, and
one can define the adjoint $S_L^{\dag}$ of the operator $S_L$ as well as the
Laplace-Beltrami operator $\{S_L\, ,S_L^{\dag}\}$, the kernel of which gives
the cohomology space of  $S_L$ \cite{[16]}.}  of the fields, sources and their
derivatives. We suppose that we have a counting operator N, with non-negative
eigenvalues $\nu = 0,1,2,...$, commuting with the Faddeev-Popov charge operator
and that decomposes the linear space V in sectors $V^{(\nu)}$ and the operator
$S_L$ in $S_L^{(\nu)}$ :
\beq\label{E1}
S_L = \sum_{\nu = 0}^{\infty} S_L^{(\nu)} {\rm\ \ \ such \; that\ \ } [N,
S_L^{(\nu)}] = \nu S_L^{(\nu)}
\eeq
The nihilpotency of $S_L$ induces on the $ S_L^{(\nu)}$ operators the relations
:
\beq\label{E2}
\sum_{\mu = 0}^{\nu} S_L^{(\mu)}S_L^{(\nu - \mu)} = 0 \ \ \  \nu = 0,1,2,..
\eeq
hence $ S_L^{(0)}$ is still a nihilpotent operator.

Let us first analyze the {\bf Faddeev-Popov charged sectors} with the
hypothesis :
\beq\label{E3}
S_L^{(0)} \Ga = 0 \Rightarrow  \Ga = S_L^{(0)} \De \; .
\eeq
Then, the filtration of equation $S_L \Ga = 0$ first gives $S_L^{(0)} \Ga^{(0)}
= 0$ and from hypothesis (\ref{E3}) it results that $\Ga^{(0)} = S_L^{(0)}
\De^{(0)}$. The next step is :
$$S_L^{(0)} \Ga^{(1)} + S_L^{(1)} \Ga^{(0)} = 0\ \ \Rightarrow \ \  S_L^{(0)}
\Ga^{(1)} + S_L^{(1)}  S_L^{(0)} \De^{(0)} = 0$$
and, due to equ. (\ref{E2}), this implies $  S_L^{(0)} [\Ga^{(1)} - S_L^{(1)}
\De^{(0)}] = 0 \;$.
\newline $\Ga^{(1)} - S_L^{(1)} \De^{(0)}$ being in a  Faddeev-Popov charged
sector, one gets from (\ref{E3}) $\Ga^{(1)} - S_L^{(1)} \De^{(0)} =  S_L^{(0)}
\De^{(1)} \; .$ At this point, one has :
\beq\label{E44}
\Ga |_{1} = \Ga^{(0)} + \Ga^{(1)} =  [S_L^{(0)} +  S_L^{(1)}] \De^{(0)} +
S_L^{(0)} \De^{(1)} = [S_L \De]|_{1} \;\;\; {\rm ....\ e.t.c.}
\eeq
Then we have the first part of the theorem : the cohomology of $S_L$ vanishes
in the  Faddeev-Popov charged sectors.

Let us now analyze the {\bf neutral sector} with the hypothesis :
\beq\label{E4}
S_L^{(0)} \Ga = 0 \Rightarrow  \Ga = S_L^{(0)} \De + \tilde{\De}
\eeq
Then, the filtration of equation $S_L \Ga = 0$ first gives $S_L^{(0)} \Ga^{(0)}
= 0$ and from hypothesis (\ref{E4}) it results that $\Ga^{(0)} = S_L^{(0)}
\De^{(0)} + \tilde{\De}^{(0)}$. The next step is :
$$S_L^{(0)} \Ga^{(1)} + S_L^{(1)} \Ga^{(0)} = 0 \ \ \Rightarrow \ \  S_L^{(0)}
\Ga^{(1)} + S_L^{(1)}  S_L^{(0)} \De^{(0)} + S_L^{(1)}\tilde{\De}^{(0)}= 0 $$
and, due to equ. (\ref{E2}), this implies $  S_L^{(0)} [\Ga^{(1)} - S_L^{(1)}
\De^{(0)}] =  -S_L^{(1)}\tilde{\De}^{(0)} \; .$ As a consequence of the
nihilpotency of $S_L^{(0)}$ , this gives
$ S_L^{(0)}[S_L^{(1)}\tilde{\De}^{(0)}] = 0$.
\newline The Faddeev-Popov charge of $S_L^{(1)}\tilde{\De}^{(0)}$ being +1, one
gets from (\ref{E3}) :
\beq\label{E5}
S_L^{(1)}\tilde{\De}^{(0)} = S_L^{(0)}\bar{\De}^{(1)} .
\eeq
At this point, one has : $  S_L^{(0)} [\Ga^{(1)} - S_L^{(1)} \De^{(0)}
+\bar{\De}^{(1)}] = 0\ ,$ which, according to (\ref{E4}) solves to : $\Ga^{(1)}
= S_L^{(1)} \De^{(0)} - \bar{\De}^{(1)} + S_L^{(0)} \De^{(1)} +
\tilde{\De}^{(1)} \;\;.$

\noindent This finaly gives :
\beqa\label{E6}
\Ga |_{1} & = & \Ga^{(0)} + \Ga^{(1)} =  [S_L^{(0)} +  S_L^{(1)}] \De^{(0)} +
S_L^{(0)} \De^{(1)} + \tilde{\De}^{(0)} + \tilde{\De}^{(1)} - \bar{\De}^{(1)}
\nnb \\
& = & [S_L \De]|_{1} + \tilde{\De}|_{1} - \bar{\De}|_{1} \;\;\; {\rm
....e.t.c.}
\eeqa
where $ \bar{\De}^{(1)}$, being determined by (\ref{E5}) up to $ S_L^{(0)}
\bar{\bar{ \De}}^{(1)}$ which would unessentially modify $\De$, adds {\bf no
new parameter } to the cohomology $\tilde{\De}$ of $ S_L^{(0)}$.

What remains to show is that the cohomology space of  $ S_L$ is not smaller
that the one of  $ S_L^{(0)}$ in that neutral sector.
Supposes that there exists some $\Ga$ belonging to the cohomology space of $
S_L^{(0)} $ :
$$ S_L^{(0)} \Ga = 0 {\rm \ \ \  with \ } \Ga\neq  S_L^{(0)} \tilde{\De} \; .$$

\noindent The previous demonstration then built $\Ga '$, a cocycle of $S_L$ and
let us suppose it to be a coboundary of $S_L$ :
\beq\label{E7}
S_L\Ga ' = 0  {\rm \ \ \ and \ \ } \Ga ' = S_L\tilde{\De} '\;\; .
\eeq
At the lowest order this would give a contradiction as $\Ga^{(0)} \equiv
\Ga'^{(0)} $. Then $\Ga^{(0)} = 0 =  S_L^{(0)} \tilde{\De '}^{(0)} $. At the
next order - in fact the first non-vanishing one -, one gets :
\beq\label{E8}
\Ga^{(1)} \equiv \Ga'^{(1)}  = S_L^{(0)} \tilde{\De '}^{(1)} + S_L^{(1)}
\tilde{\De '}^{(0)}
\eeq
But $S_L^{(0)} \tilde{\De '}^{(0)} = 0 $ where $\tilde{\De '}^{(0)}$ belongs to
the  Faddeev-Popov charge -1 sector \footnote{\ This gives a hint : if the
cohomology of $ S_L^{(0)}$ is not empty in the Faddeev-Popov charge -1 sector,
and only in that case, the cohomology spaces of $ S_L $ in the Faddeev-Popov
neutral sector may be smaller than the one of  $ S_L^{(0)}$. This is for
example the case when extra isometries do exist.} . From the hypothesis of the
theorem, this gives $\tilde{\De '}^{(0)} = S_L^{(0)} \tilde{\tilde{\De
'}}^{(0)}$ and equation (\ref{E8}) leads to :
$$\Ga^{(1)} \equiv \Ga'^{(1)}  = S_L^{(0)}[ \tilde{\De '}^{(1)} -  S_L^{(1)}
\tilde{\tilde{\De '}}^{(0)}] \ .$$
This means that $\Ga$ would be a coboundary of $ S_L^{(0)}$ (at this order $\Ga
\equiv \Ga^{(1)} $). The contradiction then gives again $\Ga^{(1)} = 0 $
....e.t.c. Then we have the announced isomorphism between the cohomology spaces
of   $ S_L $ and $ S_L^{(0)} $ in the Faddeev-Popov neutral sector.

This schematic proof illustrates the fact that a non vanishing cohomology for
$S_L^{(0)}$ in a given Faddeev-Popov charge $\nu$ sector $a\ priori$ obstructs
the construction of the cocycles of $ S_L$ in the Faddeev-Popov charge $\nu$-1
and transforms into coboundaries of $ S_L$ some of the cohomology elements of $
S_L^0$ in the Faddeev-Popov charge $\nu$+1 (see the appendix for some details).

This ends the, schematic, proof of the filtration theorem (for a complete
proof, see \cite{[13]},\cite{[7]}).

$$ $$
We shall now apply our methods to N=1 supersymmetric non-linear $\si$ models in
components fields.

\section{N=1 supersymmetric non-linear $\si$ models}
Equations (\ref{B6},\ref{B8}) respectively define the classical action and the
nihilpotent linearized Slavnov operator.
$N_{field}$ being the operator counting the number of fields (and of their
derivatives) :
\beq\label{F11}
N_{field}\De \equiv \left[\int d^2x \phi(x)\frac{\de}{\de \phi(x)} \right] \De
\;\; ,
\eeq
we will take as ghost number preserving counting (filtration) operator :
\beqa\label{F2}
N &=& N_{fields, sources} + \nnb\\
& + & \sum _{fields,\; sources,\; ghosts} {\rm \left[ spin\ of\ the\ field\ ,
source \ or \ ghost \right]x}N_{field,\; source,\; ghost} \nnb \\
&\equiv &\int d^2x \{ \phi^i\frac{\de}{\de \phi^i} + {3\over2}
\psi^a_{+}\frac{\de}{\de \psi^a_{+}} + {3\over2} \psi^a_{-}\frac{\de}{\de
\psi^a_{-}}
+ \eta_i\frac{\de}{\de \eta_i} + {3\over2} \ga_a^{+}\frac{\de}{\de \ga_a^{+}} +
{3\over2} \ga_a^{-}\frac{\de}{\de \ga_a^{-}} \} \nnb \\
 & + & {1\over2}( C^+\frac{\de}{\de C^+} + C^-\frac{\de}{\de C^-}) \ \ .
\eeqa
$e^a_{i}(0)$ being the vielbein $e_a^{i}(\phi)$ at $\phi \equiv 0, \ S_L^{0}$
is readily obtained as :
\beqa\label{F3}
S_L^{0} & \equiv & \int d^2 x \{  \left[ iC^+e_i^{a}(0)\partial_+\phi^i
\frac{\de}{\de\psi^a_{+}} +[2i\de_{ab}\partial_-\psi^b_{+} -
C^+e_a^{i}(0)\eta_i]\frac{\de}{\de\ga_a^{+}} -
g_{ij}(0)\partial^2_{+-}\phi^j\frac{\de}{\de\eta_i}\right] \nnb \\
& + & \left[ iC^-e_i^{a}(0)\partial_-\phi^i \frac{\de}{\de\psi^a_{-}}
+[2i\de_{ab}\partial_+\psi^b_{-} - C^-e_a^{i}(0)\eta_i]\frac{\de}{\de\ga_a^{-}}
- g_{ij}(0)\partial^2_{+-}\phi^j\frac{\de}{\de\eta_i}\right] \}
\eeqa

\subsection{The cohomology of $S_{L}^{0}$}
The most general functional in the fields, sources, ghosts and their
derivatives, of a given Faddeev-Popov charge, is built using Lorentz and parity
invariance (see footnote 7) and power counting \footnote {\ The canonical
dimensions of $C^{\pm},\ \phi^i(x),\ \psi^a_{\pm}(x),\ \ga_a^{\pm}(x) $ and
$\eta_i(x)$ are respectively $-{1\over 2},\; 0,\; +{1\over 2},\; +{3\over 2}$
and +2.}.

\subsubsection{The Faddeev-Popov negatively charged sectors}
The most general functional in the fields, sources, ghosts and their
derivatives, of Faddeev-Popov charge -1 is \footnote{\ More negatively
Faddeev-Popov charged functionals do not exist.} :
\beq\label{F4}
\De_{[-1]} \equiv \int d^2 x \{ [\ga_a^{+}\psi_+^{b}
+\ga_a^{-}\psi_-^{b}]S^a_{b}(\phi) + \eta_i T^i(\phi) \}
\eeq
The condition $S_{L}^{0}\De_{[-1]} = 0$ easily gives :
$$S^a_{b}(\phi) = 0 \ ;\ \ \ \int d^2 x g_{ij}(0)\partial_{+-}^2 \phi^i
T^j(\phi) = 0 \Leftrightarrow  g_{ik}(0)T^k_{,\; j}(\phi) + g_{jk}(0)T^k_{,\;
i}(\phi) = 0\ .$$
This would mean that $T^i(\phi)$ is a Killing vector with respect to the flat
approximation $g_{ij}(0)$ of the metric $g_{ij}[\phi]$. As a matter of facts,
due to the simplicity of  $\De_{[-1]}$, the cohomology of the complete  $S_{L}$
operator in the Faddeev-Popov charge sector -1 is easily obtained, and the
vector $T^i[\phi]$ should satisfy :
$$ \int d^2x \frac{\de A^{inv.}}{\de\phi^i(x)} T^i[\phi(x)] = 0 \
\Leftrightarrow \ T^i[\phi] \ {\rm is\ a\ Killing\ vector\ for\ the\ metric\
g_{ij}[\phi]\ .}$$
Then, in the absence of Killing vectors, there are no Faddev-Popov negatively
charged cocycles - nor coboundaries, see footnote 11.

\subsubsection{The Faddeev-Popov 0 charge sector }
The most general functional in the fields, sources, ghosts and their
derivatives, of Faddeev-Popov charge 0, depends on 10 functions of $\phi$ and
these 10 monomials can be ordered with respect the total spin of the fields,
sources and ghosts composing them :
\beqa\label{F5}
\De_{[0]}^0 &\equiv & \int d^2 x \partial_+\phi^i\partial_-\phi^j
T^1_{ij}(\phi) \nnb \\
\De_{[0]}^1 &\equiv & \int d^2 x \{ [\psi_+^{a}\partial_- \psi_+^{b} +
\psi_-^{a}\partial_+ \psi_-^{b}]T^2_{ab}(\phi) +  [\psi_+^{a}
\psi_+^{b}\partial_-\phi^i +
\psi_-^{a}\psi_-^{b}\partial_+\phi^i]T^3_{abi}(\phi) + \nnb \\
& + & [\ga^+_{a} C^+\partial_+\phi^i + \ga^-_{a}
C^-\partial_-\phi^i]T^{5a}_{i}(\phi) + \eta_i[C^+\psi^a_+ + C^-\psi^a_-
]T^{8i}_{a}(\phi) \} \nnb\\
\De_{[0]}^2 &\equiv & \int d^2 x \{
[\psi_+^{a}\psi_+^{b}\psi_-^{c}\psi_-^{d}]T^4_{abcd}(\phi) + \nnb \\
& + & [\ga^+_{a} C^+\psi_+^{b} \psi_+^{c} + \ga^-_{a} C^-\psi_-^{b}
\psi_-^{c}]T^{6a}_{bc}(\phi) + [\ga^+_{a} C^-\psi_+^{b} \psi_-^{c} + \ga^-_{a}
C^+\psi_-^{b} \psi_+^{c}]T^{7a}_{bc}(\phi) +\nnb \\
& + & [\ga^+_{a}\ga^+_{b} (C^-)^2 + \ga^-_{a}\ga^-_{b} (C^+)^2]T^{9ab}(\phi) +
[\ga^+_{a}\ga^-_{b} C^+C^-]T^{10ab}(\phi) \}
\eeqa

The condition $S_{L}^{0}\De_{[0]} = 0$ can be analysed in each spin sector
separatly :

\noindent i) $\De_{[0]}^0$ is not constrained, but as coboundaries exist :
$$S_{L}^{0}\tilde{\De}_{[-1]}^0 =  S_{L}^{0} \left( \int d^2 x \eta_i T^i(\phi)
\right)\ ,$$
this means that some freedom on $ T^1_{ij}(\phi) $ corresponds to a trivial
cohomology, $i.e.$ to the expected effect on the metric of the field
redefinition freedom :
$$\phi^i \rightarrow \phi^i + T^i(\phi)\ ,$$

\noindent ii) $\De_{[0]}^1$  is  constrained, and the relations that one
obtains among the 4 functions $ T^2, \;T^3, \;T^5$ and $T^8$ give :
$$\De_{[0]}^1 \equiv S_{L}^{0}\left[ - \int d^2 x \{ [\ga_a^{+}\psi_+^{b}
+\ga_a^{-}\psi_-^{b}]e^a_{i}(0)T^{8i}_b(\phi) \} \right] .$$
This trivial cohomology corresponds to a $ \psi^a_{\pm}(x)$ non-linear field
redefinition :
$$\psi_\pm^{a} \rightarrow \psi_\pm^{a}  + e^a_{i}(0)T^{8i}_b(\phi)\psi_\pm^{b}
\ \ ,$$

\noindent iii) the cocycle condition $S_{L}^{0}\De_{[0]}^2 = 0$ enforces the
vanishing of  $T^4, \;T^6, \;T^7 \;T^9$ and $T^{10} $: there are no cocycle
(nor coboundaries) in that sub-sector.

To summarize, the cohomology  of $S_L^0$ in the Faddeev-Popov 0 charge sector
corresponds to the arbitrariness of a ``metric" $T^1_{ij}(\phi)$ . This
corresponds to the renormalisability in the space of metrics ({\sl i.e. \`a la
Friedan} \cite{[8]}) as mentionned in the Introduction.

\subsubsection{The Faddeev-Popov +1 charge sector}
In that sector, the most general functional in the fields, sources, ghosts and
their derivatives depends on 23 functions of $\phi$ and again these 23
monomials can be ordered with respect to the total spin of the fields, sources
and ghosts composing them :
\beqa\label{F51}
\De_{[+1]}^0 & \equiv & 0 \nnb \\
\De_{[+1]}^1 & \equiv & \int d^2 x \left\{ ( C^+\psi_+^{a}
[\partial_+\phi^i\partial_-\phi^j U^1_{aij} + \partial_+\partial_-\phi^i
U^2_{ai}] + C^+C^-\eta_i\partial_+\phi^jU^{3i}_{j} ) + ( + \leftrightarrow -)
\right\} \nnb\\
\De_{[+1]}^2 & \equiv & \int d^2 x \{ (C^+\psi_+^{a}
\left[\psi_+^{b}\psi_+^{c}\partial_-\phi^iV^1_{abc\;i} +
\psi_-^{b}\psi_-^{c}\partial_+\phi^iV^2_{abc\;i} +
\psi_+^{b}\partial_-\psi_+^{c}V^3_{abc} +
\psi_-^{b}\partial_+\psi_-^{c}V^4_{abc}\right] + \nnb \\
& + & (C^+)^2\left[ \{\ga_a^+\psi_+^{b}V^{5a}_{b\;i} +
\ga_a^-\psi_-^{b}V^{6a}_{b\;i}\} \partial_+\phi^i +
\ga_a^+\partial_+\psi_+^{b}V^{7a}_{b} + \ga_a^-\partial_+\psi_-^{b}V^{8a}_{b}
\right]  +\nnb \\ & + & C^+C^-\ga_a^+\left[\psi_-^{b}\partial_+\phi^i
V^{9a}_{b\;i} + \partial_+\psi_-^{b}V^{10a}_{b}\right] +\nnb \\
& + & \eta_iC^+\psi_+^{a}\left[C^+\psi_+^{b}V^{11\;i}_{ab} +
C^-\psi_-^{b}V^{12\;i}_{ab}\right] + \eta_i(C^+)^2C^-\ga_a^- V^{13a\;i} ) + (+
\leftrightarrow -) \} \nnb \\
\De_{[+1]}^3 & \equiv & \int d^2 x \{ (
C^+\psi_+^{a}\psi_+^{b}\psi_+^{c}\psi_-^{d}\psi_-^{e}W^1_{abcde} + \nnb \\
& + & C^+\psi_+^{b}\psi_+^{c}\left[C^+(\ga_a^+\psi_+^{d}W^{2a}_{bcd}+
\ga_a^-\psi_-^{d}W^{3a}_{bcd}) + C^-\ga_a^+\psi_-^{d}W^{4a}_{bcd}\right] +\nnb
\\
& + & (C^+)^2 \ga_a^-\left[\ga^-_{b} C^+\psi_+^{c}W^{5ab}_{c} + \ga^+_{b}
C^-\psi_+^{c}W^{6ab}_{c} + \ga^-_{b} C^-\psi_-^{c}W^{7ab}_{c}\right]) + (+
\leftrightarrow -) \}
\eeqa
The cocycle condition on $\De_{[+1]}$ is found to enforce relations among those
23 functions such as  $\De_{[+1]}$ finally depends only on 9 functions and may
be identified with the coboundary $S_{L}^{0}\De_{[0]}$, where $\De_{[0]}$ is
equal to the $  \De_{[0]}^1 +  \De_{[0]}^2 $ of equations (\ref{F5}). As a
consequence  $S_{L}^{0}$ has no anomaly.
$$ $$
It results from the whole subsection 4.1 that - at least in the absence of
Killing vectors for the metric - the cohomology of $S_{L}^{0}$ vanishes in the
charged Faddeev-Popov sectors and is identified in the neutral sector by a
generic symmetric metric tensor in the target space (up to an arbitrary change
of coordinates).

\subsection{The cohomology of $S_{L}$ {\sl (in the absence of Killing vectors
for the classical metric)}}
The hypothesis of our filtration theorem being satisfied, we get the desired
results for the cohomology of the whole Slavnov operator $S_{L}$ :

$\bullet$ the Slavnov identity is not anomalous [absence of Faddeev-Popov
charge +1 cohomology], which means that, as expected, N=1 supersymmetry is
renormalisable (in the 4 dimensional case this was proved in \cite {[18]}).
Notice that this property is not changed by a possible Faddeev-Popov charge -1
cohomology for $S_{L}^{0}$.

$\bullet$ moreover, the cohomology of $S_{L}$ in the Faddeev-Popov neutral
sector is identified by a generic, symmetric, metric tensor in the target space
and  may be obtained as :
$$\De_{[0]}[equ. (\ref{F5})] \equiv \Ga^{tot.}[equ. (\ref{B6})] $$ with $g_{ij}
\equiv t_{ij}$, e.t.c.
In this case of d=2, N=1 non-linear $\si$ models, the renormalisation algorithm
{\sl a priori} does not change the number of parameters with respect to the one
of the classical action\footnote{\ To be made more precise, this assertion
supposes a true definition of the classical action, for example through extra
isometries (see the next section).}. Then we have the ``stability" of the
classical action in the space of metrics, $i.e.$ the full renormalisability of
the theory.
Of course, as usual the trivial cohomology
$S_L\De_{[-1]}(T^i[\phi],S^a_b[\phi])$ corresponds to the non-linear fields and
sources reparametrisations, here according to :
\beqa\label{F44}
\phi^i \rightarrow \phi^i + T^i(\phi)\ & , &\ \eta_i \rightarrow \eta_i -
\eta_kT^k_{,\; i} -  S^a_{b\;,i}(\phi)[\ga^+_{a}\psi_+^{b} +\ga^-_{a}\psi_-^{b}
]  \nnb\\
\psi_\pm^{a} \rightarrow \psi_\pm^{a}  - S^a_b(\phi)\psi_\pm^{b} \ & ,& \
\ga^\pm_{a} \rightarrow \ga^\pm_{a} +  S^b_{a}(\phi)\ga^\pm_{b}
\eeqa

\section{Cohomology of supersymmetry in the presence of Killing vectors for the
classical metric :\newline {\sl the example of coset-spaces}}
As previously mentionned, the existence of Killing vectors is responsible for a
non vanishing cohomology in the Faddeev-Popov charge -1 sector, which will
restrict the one in the Faddeev-Popov 0 charge sector (the Action). As a matter
of facts, the occurence of Killing vectors reveals the presence of extra
symmetries whose renormalisability has also to be studied, from the very
beginning, as these isometries are needed for a precise definition of the
classical action and of the true physical parameters. As an example, in this
section we extend to N=1 supersymmetric models the results of our previous
analysis on the renormalisability of bosonic non-linear $\si$ models built on
compact homogeneous spaces G/H \cite{[7]}.

The non-linear isometries may be writen as :
\beqa\label{g1}
\de\phi^i &=& \epsilon^{j}W^{i}_{j}[\phi] \nnb \\
\de\psi^{i}_{\pm} &=& \epsilon^{j}W^{i}_{j,k}[\phi]\psi^{k}_{\pm}
\eeqa
where the $W^i_j$ \ , j = 1,2,..n are n  Killing vectors for the metric $
g_{ij}$ :
$$g_{il}\nabla_j W^l_k + g_{jl}\nabla_i W^l_k  = 0 \ ,$$
and where $\nabla_i$ is the covariant derivative with (torsionless) symmetric
connexion  associated to the metric $ g_{ij}\ .$

The homogeneity of the coset space G/H means that :
\beq\label{g11}
W^i_j [\phi] = \delta^i_j + ...,
\eeq

With regard to the linear isometries, we know from  the appendix A of
\cite{[7]} that they cause no difficulty in the renormalisation program, at
least when the corresponding Lie group H is a compact one. In the following, we
then restrict ourselves to H-invariant integrated local functionals in the
fields, sources, ghosts and their derivatives.

Using tangent space fermions as defined in subsection 2.2, transformations
(\ref{g1}) write :
\beqa\label{g2}
\de\phi^i &=& \epsilon^{j} e^{i}_{a}W^{a}_{j} \nnb \\
\de\psi^{a}_{\pm} &=&  \epsilon^{j}[\nabla_c W^a_{j} -
\om^a_{bc}W^b_j]\psi^{c}_{\pm}\ .
\eeqa
Of course, the supersymmetry transformations (\ref{B4}) commute with the
previous ones (\ref{g2}) and the commutation relations of the non-linear
transformations (\ref{g1}) being those of a standard Lie algebra,
equ.(\ref{a2}) still specifies to equ.(\ref{B5}). One then introduce constant
anticommuting Faddeev-Popov ghosts $C^j$ \footnote{\ Due to the previous
remark, and  in the same manner as we did before, not introducing ghosts for
the translations, it is not necessary to add ghosts associated to the linear
isometries.}(of vanishing canonical dimension), and modify the total effective
action of equ.(\ref{B6}):
\beqa\label{g6}
\Ga^{tot.} & = & A^{inv.} + \int d^2 x \{\ga_{a}^{+}(x)\left(C^{j}[\nabla_c
W^a_{j} - \om^a_{bc}W^b_j]\psi^{c}_{+} + iC^+ e^{a}_{+} +
\om^a_{bc}\psi^{c}_{+}(C^+\psi^{b}_{+} + C^-\psi^{b}_{-})\right)\nnb \\
& + & \ga_{a}^{-}(x)\left(C^{j}[\nabla_c W^a_{j} - \om^a_{bc}W^b_j]\psi^{c}_{-}
+ iC^- e^{a}_{-} + \om^a_{bc}\psi^{c}_{-}(C^+\psi^{b}_{+} +
C^-\psi^{b}_{-})\right) \nnb \\
& + & \eta_i(x)( C^j e^{i}_{a}W^{a}_{j} + C^+ e^{i}_{a}\psi^{a}_{+} + C^-
e^{i}_{a}\psi^{a}_{-} )\} \nnb \\
& + & {1\over4}\int d^2 x \de^{ab}\{\ga_{a}^{+}(x)\ga_{b}^{+}(x)(C^-)^2
+\ga_{a}^{-}(x)\ga_{b}^{-}(x)(C^+)^2 - 2\ga_{a}^{+}(x)\ga_{b}^{-}(x)(C^+ C^-)\}
\eeqa
The Slavnov identity of equ.(\ref{a10}) still holds and the linearized Slavnov
operator (\ref{B8}) keeps the same structure :
\beqa\label{g8}
S_L \equiv \int & d^2 x & \{ \left( \frac{\de \Ga^{tot.}}{\de \ga_a^{+}}\right)
\frac{\de}{\de \psi^a_{+}} + \left( \frac{\de \Ga^{tot.}}{\de \ga_a^{-}}\right)
\frac{\de}{\de \psi^a_{-}} + \left( \frac{\de \Ga^{tot.}}{\de
\psi^a_{+}}\right) \frac{\de}{\de \ga_a^{+}}
+ \left(\frac{\de \Ga^{tot.}}{\de \psi^a_{-}}\right) \frac{\de}{\de \ga_a^{-}}
\nnb \\
& + &  \left(  \frac{\de \Ga^{tot.}}{\de \eta_i}\right) \frac{\de}{\de \phi^i}
+ \left(\frac{\de \Ga^{tot.}}{\de \phi^i}\right) \frac{\de}{\de \eta_i} \}
\eeqa
and is nihilpotent : $(S_L)^2 = 0$, when acting in the space of H-invariant
integrated local functionals. For further use, and thanks to the commutation of
transformations (\ref{g2}) and (\ref{B4}), we split  $S_L$ in :
$$S_L = S_L^s + S_L^w \ ,$$
where $S_L^w$ is the nihilpotent Slavnov operator associated to the homogenous
transformations (\ref{g2}) and is linear in $C^j$ and we have :
$$(S_L^s)^2 = (S_L^w)^2 = S_L^w S_L^s + S_L^s S_L^w = 0\ .$$

The classical action and the nihilpotent linearized Slavnov operator being so
defined, we add to the grading operator (\ref{F2})  the operator counting the
number of ghosts $C^j\ .\ S_L^{0}$ is readily obtained as  $ S_L^{s|0} +
S_L^{w|0}$ where $S_L^{s|0}$ was given in equ.(\ref{F3}) and :
$$S_L^{w|0} = C^i \int d^2 x\frac{\de}{\de\phi^i(x)} $$
where the fundamental homogeneity property  (\ref{g11}) has been used.

We are now in a position to analyse the cohomology of $ S_L^{0}$, using the
results of subsection 4.1 for $ S_L^{s|0}$ and of \cite{[7]} for $S_L^{w|0}\ .$
\subsection{The cohomology of $S_L^0$}

\subsubsection{The Faddeev-Popov negatively charged sectors}
The most general functional in the fields, sources, ghosts and their
derivatives, of Faddeev-Popov charge -1 is still given by equ.(\ref{F4}).
The condition $S_{L}^{0}\De_{[-1]} = 0$ easily gives :
$$S^a_{b}(\phi) = 0 \ ;\ \ \ T^i(\phi) = {\rm constant}\ .$$
Moreover, this cohomology should be H-invariant. This occurs only when among
the non-linear transformations $W_j$, there exists some, labelled by indices
$\underline{j}$, $W_{\underline{j}}$, corresponding to a subgroup X of G that
commutes with H. As explained in detail in \cite{[7]}, in such a case, some of
the parameters that define the invariant action $\int d^2x
g_{ij}[\phi]\partial_+\phi^i\partial_-\phi^j$ becomes unphysical ones. This
corresponds to the fact that the non vanishing cohomology $T^{\underline{i}}$
in the Faddeev-Popov -1 charge sector restricts the cohomology of $S_L$ in the
Faddeev-Popov 0 charge sector (see the appendix and \cite{[b10]}).

\subsubsection{The Faddeev-Popov 0 charge sector}
The most general functional in the fields, sources, ghosts and their
derivatives, of Faddeev-Popov charge 0 may be split into two parts :
$\De_{[0]}^0$ and $\De_{[0]}^1$ according to their number of ghosts $C^i.\
\De_{[0]}^0 $ is given by (\ref{F5}) and
$$\De_{[0]}^1 = C^j\int d^2 x \{ [\ga_a^{+}\psi_+^{b}
+\ga_a^{-}\psi_-^{b}]S^a_{bj}(\phi) + \eta_i T^i_j(\phi) \}\ \equiv
C^j\De_{[-1]\;j}^0$$
where $\De_{[-1]\;j}^0$ has the same expression as $\De_{[-1]}$ of (\ref{F4})
with tensors having one more index j.
The condition $S_{L}^{0}\De_{[0]} = 0$ is then analysed into 3 steps
corresponding to the total $C^j$ ghost number :
$$\bullet  S_L^{s|0} \De_{[0]}^0 = 0 \ \
\stackrel{subsec.(4.1.2)}{\Leftrightarrow}\\
 \De_{[0]}^0 = \int d^2x \partial_+\phi^i\partial_-\phi^j T_{ij}[\phi] +
S_L^{s|0}\De_{[-1]}[S^a_{1b},\ T^i_1]$$ where the symmetric tensor $T_{ij}$ is,
for the moment, unconstrained.
$$\bullet  S_L^{w|0} \De_{[0]}^1 = 0 \ \Leftrightarrow \ \De_{[0]}^1 =
S_L^{w|0} \De_{[-1]}[S^a_{2b},\ T^i_2] $$
$$\bullet  S_L^{w|0} \De_{[0]}^0 + S_L^{s|0} \De_{[0]}^1 = 0 \\ \Leftrightarrow
\ S_L^{s|0}S_L^{w|0} \De_{[-1]}[S^a_{1b} - S^a_{2b},\ T^i_1 - T^i_2] =
S_L^{w|0}\int d^2x \partial_+\phi^i\partial_-\phi^j T_{ij}[\phi]$$
$$\Rightarrow \ S^a_{1b} - S^a_{2b} = {\rm constant}\ ,\ T_{ij}[\phi] =
-(g_{il}(0)[T^l_1 - T^l_2]_{,j} + g_{jl}(0)[T^l_1 - T^l_2]_{,i}) + \la_{ij} $$
where $\la_{ij}$ is a constant H-invariant tensor.

Finally this gives :
\beq\label{h1}
\De_{[0]} = \la_{ij}\int d^2x \partial_+\phi^i\partial_-\phi^j +
S_{L}^{0}\De_{[-1]}[S^a_{1b}, T^i_2]
\eeq
This means that, as in the purely bosonic case \cite{[7]}, the dimension of the
cohomology space of $S_L^{0}$ in the Faddeev-Popov neutral sector is given by
the number of H-invariant symmetric 2-tensors.

\subsubsection{The Faddeev-Popov +1 charge sector}
The most general functional in the fields, sources, ghosts and their
derivatives, of Faddeev-Popov charge +1 may be split into three parts :
$\De_{[+1]}^0,\ \De_{[+1]}^1$ and $\De_{[+1]}^2$ according to their number of
ghosts $C^i\ .\ \De_{[+1]}^0 $ is given by (\ref{F51}), $\De_{[+1]}^1 \equiv
C^j \De_{[0]j}^0$ where $\De_{[0]j}^0$ has the same expression as $\De_{[0]}$
given by (\ref{F5}) with tensors having one more index j, and $\De_{[+1]}^2
\equiv C^jC^k \De_{[-1]jk}^0$ where $\De_{[-1]jk}^0$ has the same expression as
$\De_{[-1]}$ given by (\ref{F4}) with tensors having two more indices j and k.
The condition $S_{L}^{0}\De_{[+1]} = 0$ is then analysed into 4 steps
corresponding to the total $C^j$ ghost number :
$$\bullet  S_L^{s|0} \De_{[+1]}^0 = 0 \ \
\stackrel{subsec.(4.1.3)}{\Leftrightarrow}\ \ \De_{[+1]}^0 = S_L^{s|0}
\De_{[0]}^0 $$
$$\bullet  S_L^{w|0} \De_{[+1]}^2 = 0 \ \Leftrightarrow \ \De_{[+1]}^2 =
S_L^{w|0} \De_{[0]}^1 = -C^k S_L^{w|0} \De_{[-1]k}^0 [S_{1},\ T_{1}] $$
$$\bullet  S_L^{w|0} \De_{[+1]}^0 + S_L^{s|0} \De_{[+1]}^1 = 0 \\
\Leftrightarrow \ -C^k S_L^{s|0}\left[\De_{[0]k}^0 - \int d^2x
\frac{\de}{\de\phi^k(x)}\De_{[0]}^0 \right] = 0 $$
$$\stackrel{subsec.(4.1.2)}{\Leftrightarrow}\ \ \De_{[0]k}^0 = \int d^2x
T^1_{ijk}\partial_+\phi^i\partial_-\phi^j -  S_L^{s|0}\De_{[-1]k}^0 [S_{2},\
T_{2}] + \int d^2x\frac{\de}{\de\phi^k(x)}\De_{[0]}^0 $$
$$\Rightarrow \ \De_{[+1]}^1 = C^k \int d^2x
T^1_{ijk}\partial_+\phi^i\partial_-\phi^j +  S_L^{s|0}\De_{[0]}^1 [S_{2},\
T_{2}] + S_L^{w|0} \De_{[0]}^0 $$

$$\bullet  S_L^{w|0} \De_{[+1]}^1 + S_L^{s|0} \De_{[+1]}^2 = 0 \ \
\Leftrightarrow \ \ C^k S_L^{w|0}\int d^2x
T^1_{ijk}\partial_+\phi^i\partial_-\phi^j = C^k S_L^{s|0}
S_L^{w|0}\De_{[-1]k}^0 [S_{1}-S_{2},\ T_{1}-T_{2}] $$
$$\Rightarrow \ [S_{1}-S_{2}]^a_{bk} = \partial_k [S_{1}-S_{2}]^a_{b}\ ;\
T^1_{ijk} = \partial_k T_{ij}[\phi] - (g_{il}(0)[T^l_{1k} - T^l_{2k}]_{,j} +
g_{jl}(0)[T^l_{1k} - T^l_{2k}]_{,i})$$

Finally this gives :
\beq\label{h2}
\De_{[+1]} = S_{L}^{0}\left[ \De_{[0]}^0 + \De_{[0]}^1 [S^a_{2b},\ T^i_{1}] +
\int d^2x T_{ij}[\phi]\partial_+\phi^i\partial_-\phi^j \right] =
S_{L}^{0}\De_{[0]}
\eeq
This means that the cohomology of $S_{L}^{0}$ in the Faddeev-Popov charge 1
sector vanishes.

\subsection{The cohomology of $S_{L}$}
It results from the whole subsection 5.1 that the cohomology of $S_{L}^{0}$
vanishes in the charged Faddeev-Popov sectors and is identified in the neutral
sector by a generic H-invariant constant symmetric 2-tensor. Then, the
hypothesis of our filtration theorem being satisfied, we get the desired
results for the cohomology of the whole Slavnov operator $S_{L}\ :$

\noindent $\bullet$ the Slavnov identity is not anomalous [absence of
Faddeev-Popov charge +1 cohomology],

\noindent $\bullet$ the cohomology of $S_{L}$ in the Faddeev-Popov neutral
sector is, as the classical action, identified by by a generic \footnote{\ up
to some Faddeev-Popov charge -1 cohomology, which restricts the number of
physical parameters of the classical action (subsection (5.1.1) and appendix
7.2).} H-invariant constant symmetric 2-tensor [stability of the theory].

This means that, as expected, N=1 supersymmetric non-linear $\si$ models built
on compact homogeneous spaces are renormalisable.

\section{Concluding remarks}

We have analysed in a regularisation free manner the all-order
renormalisability of N=1 supersymmetric non-linear $\si$ models in component
fields. Using a conveniently chosen gradation, we proved the absence of
supersymmetry anomaly and the renormalisability of the theory ``in the space of
metrics". For the special class of $\si$ models built on an homogeneous
manifold (usual non linear $\si$ models on coset space), our work extends the
renormalisability proof given for the bosonic case in ref.\cite{[7]} to the N=1
supersymmetric case (up to infra-red analysis).

The quantization of the N=2 and 4 supersymmetric non-linear $\si$ models of
subsection 2.3 will be studied in a second paper of this series \cite{[12]}. In
particular, the rigorous proof of the renormalisability of N=1 supersymmetric
non-linear $\si$ models in component fields given here will allow us to analyse
the extended supersymmetries in a N=1 superfield formalism (see also
\cite{[2]}).

\vfill {\bf Aknowledgements :} It is a pleasure to thank F. Delduc for his help
in the formulation of subsections 2.2 and 2.3 and for useful discussions.

\section{Appendix : the filtration theorem in the presence of a non trivial
cohomology for $S_L^{(0)}$ in the Faddeev-Popov charged sectors. }

This appendix intends to give simple proofs of the results of the original
papers (\cite{[13]},\cite{[7]} and \cite{[b10]}) in order to illustrate the
fact that a non vanishing cohomology for $S_L^{(0)}$ in a given Faddeev-Popov
charge $\nu$ sector $a\ priori$ obstructs the construction of the cocycles of $
S_L$ in the Faddeev-Popov charge $\nu$-1 and transforms into coboundaries of $
S_L$ some of the cohomology elements of $ S_L^0$ in the Faddeev-Popov charge
$\nu$+1.

\subsection{Presence of some cohomology in the Faddeev-Popov charge +1 sector.
}
As can be seen from the sketch of the proof given in subsection 3.2, this may
prevent from constructing  Faddeev-Popov 0 charge cocycles of $S_L$ starting
from those of $S_L^{(0)} .$ Indeed, suppose that there is some cohomology
begining at the filtration level $\nu$ :
\beq\label{ap1}
S_L^{(0)}\De_{[1]} = 0 \ \ \Rightarrow \ \De_{[1]} = \De_{[1]}^{an.(\nu)} +
S_L^{(0)}\De_{[0]}\ .
\eeq
The construction described through equations (\ref{E4}) to (\ref{E6}) builds
cocycles in the Faddeev-Popov 0 charge sector
$$S_L\Ga |_{(\nu -1)} = 0\ \ \Rightarrow \ \Ga|_{(\nu -1)} =
\De^{an.}_{[0]}|_{(\nu -1)} - \bar{\De}_{[0]}|_{(\nu -1)} +
\left(S_L\De_{[-1]}\right)|_{(\nu -1)}. $$
At the next order one has :
$$S_L^{(0)}\Ga^{(\nu)} + S_L^{(1)}\Ga^{(\nu -1)} + .. + S_L^{(\nu)}\Ga^{(0)} =
0$$
$$\Rightarrow \ S_L^{(0)}[\Ga^{(\nu)} - S_L^{(1)}\De_{[-1]}^{(\nu -1)} + ..] +
\{S_L^{(1)}\De_{[0]}^{an.(\nu -1)} - S_L^{(1)}\bar{\De}_{[0]}^{(\nu -1)} +...\}
= 0 .$$
{}From $S_L^{(0)}\{S_L^{(1)}\De_{[0]}^{an.(\nu -1)} -
S_L^{(1)}\bar{\De}_{[0]}^{(\nu -1)} + ... \} = 0$, the hypothesis (\ref{ap1})
gives :
\beq\label{ap2}
\{S_L^{(1)}\De_{[0]}^{an.(\nu -1)} - S_L^{(1)}\bar{\De}_{[0]}^{(\nu -1)} + ...
\} =  \De_{[1]}^{an.(\nu)} + S_L^{(0)}\bar{\De}_{[0]}^{(\nu)}
\eeq
and finally :
$$ S_L^{(0)}\left[\Ga^{(\nu)} - S_L^{(1)}\De_{[-1]}^{(\nu -1)} +...+
\bar{\De}_{[0]}^{(\nu)}\right]  + \De_{[1]}^{an.(\nu)} = 0\ ,$$
which is self contradictory, \underline{except if the coefficients} in the
$\De_{[0]}^{an.(\nu -p)}, \bar{\De}_{[0]}^{(\nu -p)}$ involved in
equ.(\ref{ap2}) are related in such a way that $\De_{[1]}^{an.(\nu)}$ does not
appear. In that case, one gets
$$\Ga^{(\nu)} = S_L^{(1)}\De_{[-1]}^{(\nu -1)} + ...  + \De_{[0]}^{an.(\nu)} -
\bar{\De}_{[0]}^{(\nu)} + S_L^{(0)}\De_{[-1]}^{(\nu)}$$
and finally
$$\Ga |_{(\nu)} = \De^{an.}_{[0]}|_{(\nu)}  - \bar{\De}_{[0]}|_{(\nu)} +
\left(S_L\De_{[-1]}\right)|_{(\nu)}  \ \ \ \ \  Q.E.D.$$
Thus, in the case of a non vanishing intersection between the cohomology in the
 Faddeev-Popov charge 1 sector and the successive images through $S_L^{(1)}$,
$S_L^{(2)}$... of the insertions $\De_{[0]}^{an.(\nu)},
\bar{\De}_{[0]}^{(\nu)}$ of Faddeev-Popov 0 charge, the aforementionned
relations reduce the number of cocycles of $S_L$ in the Faddeev-Popov 0 charge
sector with respect to the ones of $S_L^{(0)}$.

This analysis will be useful for the study of N=2 and 4 supersymmetric
non-linear $\si$ models \cite{[12]}.

\subsection{Presence of some cohomology in the Faddeev-Popov charge -1 sector.
}
As indicated in subsection 3.2, this will reduce the dimension of the
cohomology space of $S_L$ in the Faddeev-Popov 0 charge sector with respect to
the one of $S_L^{(0)}$. A complete analysis is given in  the appendix C of
ref.\cite{[b10]} where it is shown that the cohomology of $S_L$ in the
Faddeev-Popov 0 charge sector is isomorphic to the repeated quotient of the
cohomology of $S_L^{(0)}$ in the same Faddeev-Popov sector by the successive
images through  $S_L^{(1)}$, $S_L^{(2)}$... of the cohomology of $S_L^{(0)}$ in
the Faddeev-Popov -1 charge sector. This reduction is now due to the fact that
some cocycles for $S_L$ built according to equations (\ref{E4}) to (\ref{E6})
may be coboundaries when there is some $S_L^{(0)}$ cohomology in the
Faddeev-Popov -1 charge sector. For self-containtness, we give now some hints
on this mechanism.

Suppose for example that there is some cohomology in the Faddeev-Popov -1
charge sector beginning at the filtration level $\nu$ :
\beq\label{ap13}
S_L^{(0)}\De_{[-1]}^{an.(\nu)} = 0 \ \ {\rm with} \ \De_{[-1]}^{an.(\nu)} \neq
S_L^{(0)}\De_{[-2]}^{(\nu)}
\eeq
and consider $S_L^{(1)}\De_{[-1]}^{an.(\nu)}$. It is a $S_L^{(0)}$ cocycle in
the Faddeev-Popov 0 charge sector and then may be writen as :
$$ S_L^{(1)}\De_{[-1]}^{an.(\nu)} = S_L^{(0)}\tilde\De_{[-1]}^{(\nu +1)} +
\De_{[0]}^{an.(\nu +1)}\ .$$
If this really occurs, $i.e.$ if there is a non empty intersection between the
cohomology of $S_L^{(0)}$ in the Faddeev-Popov 0 charge sector and the image
through $ S_L^{(1)}$ of the one in the Faddeev-Popov -1 charge, this particular
cohomology trivializes itself. Indeed
$$\De_{[0]}^{an.(\nu +1)} = S_L^{(1)}\De_{[-1]}^{an.(\nu)} -
S_L^{(0)}\tilde{\De}_{[-1]}^{(\nu +1)} \equiv (S_L^{(0)} +
S_L^{(1)})\left(\De_{[-1]}^{an.(\nu)} - \tilde{\De}_{[-1]}^{(\nu +1)}\right) +
{\cal O} (\bar{\bar{\De}}_{[0]}^{(\nu+2)})\ ,$$
and so on on higher orders. This in particular occurred in subsections (5.1.1)
and (5.1.2) where
$$ \De_{[-1]}^{an.}= T^{\underline{i}}\int d^2x \eta_{\underline{i}} \equiv
\De_{[-1]}^{an.(1)}\ \ ,\ \ \De_{[0]}^{an.}= \la_{ij}\int d^2x
\partial_+\phi^i\partial_-\phi^j \equiv \De_{[0]}^{an.(2)}\ .$$
Then, using the invariance under (\ref{g1}) of the classical action, one may
check that :
$$S_L^{(1)}\De_{[-1]}^{an.(1)}= T^{\underline{i}}\int d^2x
\frac{\de\Ga^{tot.}}{\de\phi^{\underline{i}}}|_2 = T^{\underline{k}}
g_{ij,{\underline{k}}}(0)\int d^2x \partial_+\phi^i\partial_-\phi^j +
T^{\underline{k}}C^i W^j_{i,\,{\underline{k}}}(0)\int d^2x \eta_j $$
$$ = S_L^0\left[ -T^{\underline{k}}W^j_{i,\,{\underline{k}}}(0)\int d^2 x
\eta_j \phi^i(x) \right] +
T^{\underline{k}}\left[g_{il}(0)[W^l_{j,\,{\underline{k}}}(0) -
W^l_{\underline{k},\,j}(0)] + (i \leftrightarrow j) \right]\int d^2x
\partial_+\phi^i\partial_-\phi^j $$
does intercept $\De_{[0]}^{an.(2)}$ if $T^{\underline{i}} \neq 0 \,.$

This means that among the parameters $\la_{ij}$, the ones that are equal to
$-T^{\underline{k}}[g_{il}(0)f^l_{j{\underline{k}}} +
g_{jl}(0)f^l_{i{\underline{k}}}]$ - where $f^l_{i{\underline{k}}}$ are
structure constants of the Lie algebra of G -, are unphysical parameters as
they may be ruled out through a particular field redefinition (corresponding to
a trivial cohomology).  These  $\la_{ij}$ corresponds to X (and H)-invariant
2-tensors (see \cite{[7]} for details).

\bibliographystyle{plain}

\end{document}